\documentstyle[12pt]{article}
\begin{document}
\baselineskip 24pt
\begin{center}
\baselineskip 24pt
\noindent
{\large\bf Let the New Experiments Tell the Quantum Theory}

\vskip 24pt
\normalsize
\baselineskip 24pt
Guang-jiong Ni$^*$\footnote{$^*$ E-mail: gjni@fudan.ac.cn}\\
{\it Department of Physics, Fudan University, Shanghai, 200433, China}
\end{center}

\vskip 48pt
\baselineskip 24pt
\normalsize
\begin{center}
{\bf Abstract}
\end{center}

\leftskip 9mm
\rightskip 9mm

\vskip 12pt

\baselineskip 24pt
\indent Several new physics experiments in 1998 were performed and analyzed
\baselineskip 24pt
to show
the subtlety of quantum theory, including the ``wave-particle duality" and
the non-separability of two-particle entangled state. Here it is shown that
the measurement is bound to change the object by destroying the original
quantum coherence between the object and its environment. So the ``physical
reality" should be defined at two levels, the ``thing in itself" and the
``thing for us". The wave function in quantum mechanics is just playing the
role for connecting the two levels of matter via the fictitious measurement.


\newpage
\leftskip 0mm
\rightskip 0mm

\baselineskip 24pt
\normalsize
\indent In recent years especially in 1998, a series of important physics
\baselineskip 24pt
experiments were published. They were performed in such delicate manner and
with so amazing results that pushed the subtlety or mystery of quantum
theory in front of the physicists as well as the public in a very acute way.
These experiments can be sorted into four categories:

(1) The experimental discovery of fractional quantum Hall effect (FQHE) [1]
and the observation of fractional charge ($e/3$) in FQHE system [2,3].

(2) Direct test of wave-particle duality (complementarity) by a ``which-way"
experiment in an atom interferometer [4].

(3) Einstein-Podolsky-Rosen (EPR) experiments were performed in two-photon
entangled state to show the violation of Bell inequality under strict
Einstein locality conditions [5] or to show the quantum correlation over
long distance ($>10$km) [6]. Also an EPR experiment was achieved
at CERN to test the non-separability of entangled neutral-kaon wave function
[7].

(4) First direct observation of time-reversal non-invariance in the
neutral-kaon system [8].

In this paper we will concentrate on the (1)---(3) experiments especially (2)
and (3) because they are directly related to the fundamental interpretation
of quantum mechanics (QM). We will show that the outcome of these
experiments strongly support the validity and completeness of QM including
the Heisenberg's position-momentum uncertainty relation. However, the
correct interpretation of QM does need a clarification of an important point
of view that a quantum state before the measurement has no any information.
The latter is created only during the measurement by the object and subject
in common.

\vskip 24pt
\noindent
{\bf I. The essence of measurement}

The ``which way" (WW) experiment in an atom interferometer was proposed by
Scully {\it et al} [9] and successfully realized by DNR [4]. It is a new
realization of ideal experiment long considered by physicists , say, by
Feynman [10]. In Feynman's double-slit interference experiment of electron,
in order to determine which slit the electron goes through, a light source
is put just behind the double-slit for watching the electron. Feynman
predicted that once the light source is switched on, the interference
pattern will be washed out. And it was explained by the impact of photon on
the electron. The momentum transfer $\Delta p_x$ and the position uncertainty
of electron along the screen direction, $\Delta x$, will be constrained to
the Heisenberg's uncertainty relation:
\begin{equation}
\Delta x\Delta p_x\geq\hbar/2.
\end{equation}

As Feynman quoted from Heisenberg that ``It is impossible to design an
apparatus to determine which hole the electron passes through, that will not
at the same time disturb the electrons enough to destroy the interference
pattern." Feynman said: ``The uncertainty principle `protects' quantum
mechanics." See also some interesting discussion on this problem in Refs.
[11-13].

Instead of electrons, the $^{85}Rb$ atoms are used in DNR's experiment. The
``double-slit" is realized by the Bragg diffraction of atomic beam on two
standing light waves. The WW information is provided by two microwave pulses
which divide the splitting beams into pure internal states, either
$|2\rangle$ or $|3\rangle$ being
two hyperfine states with total angular momentum $F=2$ or 3. Once this is
done, the interference fringes are lost due to the $\langle 3|2\rangle=0$.
The analysis shows
that the momentum transfer from microwave pulse to the atom is so negligible
that it plays no role in the loss of interference. Then the authors
concluded that complementarity is not enforced by the uncertainty relation,
Eq. (1).

Undoubtedly the DNR experiment is very important for it further reveals the
essence of measurement which can be summarized as three propositions: 

(a) The measurement is bound to change the state of object.

(b) The measurement is also quantum in essence. The quantum correlation
(i.e. entanglement) between the apparatus and the object is bound to destroy
the quantum correlation (quantum coherence) originally existing in the
object.

(c) There is no any information (experimental data) existed before the
measurement is made.

Let us first look at the measurement in classical physics. The specific heat
at constant volume of certain substance is defined as
\begin{equation}
C_V=\frac{\partial U}{\partial T}\mid_V.
\end{equation}
While that at constant pressure is 
\begin{equation}
C_P=T\frac{\partial S}{\partial T}\mid_p.
\end{equation}

Note that, however, the change of the internal energy $U$, or the temperature
$T$, or the entropy $S$, even is infinitesimal but certainly can not be zero
because of the energy quantum $h\mu$ with the Planck constant $h\neq 0$.
Therefore, the
definition and value of $C_V$ or $C_P$ are endowed by the operation during the
measuring process as shown by Eq. (2) or (3) respectively. Either $C_V$ or
$C_P$ is objective in the sense of its unique value given by the measurement.
However, the simultaneous measurement of $C_V$ and $C_P$ will disturb each
other, leading to inaccuracy in the result. The reason is as follows [14]:

A measurement is always an operation procedure (denoted by $A$) for changing
the object to pick out the corresponding data (denoted by $a$):
$A\longrightarrow a$. Similarly, another measuring procedure $B$ leads to $b$:
$B\longrightarrow b$. If $A$ and $B$ are not in conformity but are imposed
simultaneously on an object, then $A$ (or $B$) would become the disturbance to
$b$ (or $a$), as shown in Fig. 1.

The great merit of quantum mechanics (QM) lies in the fact that it unveils
the truth of epistemology at the basic level. For example, the plane wave
function of a freely moving particle reads
\begin{equation}
\Psi(x,t)\sim exp\left\{\frac{i}{\hbar}(p_xx-Et)\right\}.
\end{equation}

Notice that, however, where the momentum parameter $p_x$ inside is not an
observable, i.e., not a quantity existing before the measurement. Only
during a ``space translational operation" is imposed on the wave function
as follows
\begin{equation}
-i\hbar\frac{\partial}{\partial x}\Psi
=-i\hbar \lim\limits_{\Delta x\rightarrow 0}
\frac{\Psi(x+\Delta x, t)-\Psi(x, t)}{\Delta x}=p_x\Psi
\end{equation}
can we pick out the observable $p_x$ on the right side. So the fact that in QM
a classical dynamical variable, say $p_x$, becomes an operator
\begin{equation}
\hat{p}_x=-i\hbar\frac{\partial}{\partial x}
\end{equation}
is nothing but a mathematical statement (6) of the principle of epistemology
that the measurement is always a change on the object. 

It is because the measurement of position $x$ and $p_x$ are different, they have
the meaning of complementarity. On the other hand, both $x$ and $p_x$ are related
to the measurement in the space, i.e., they have ``identity" in essence, so
the ``repulsiveness" (conflict) contained in their difference will display
inevitably under certain condition. Therefore, it seems to us that the
uncertainty relation and the complementarity just are two aspects reflecting
the same essence. There is no question about which one of them is more
fundamental.

The DNR experiment can also be explained by the general scheme with Fig. 1.
The experiment arrangement without two microwave pulses is denoted by ``$A$",
while ``$a$" denotes the appearance of interference fringe. On the other
hand, ``$B$" denotes the operation of imposing two microwave pulses for
measuring the WW information, while ``$b$" denotes the disappearance of
fringe. The repulsiveness or complementarity is contained in one expression
that the distinguishability (of WW information) $D$ and the fringe visibility
$V$ is limited by the duality relation $D^2+V^2\leq 1$ [4].

The unique feature of DNR experiment lies in the fact that they change the
internal quantum state of atom. In some sense the measurement destroys the
quantum coherence of internal state. It has no classical correspondence.

\vskip 24pt
\noindent
{\bf II. The EPR experiments and the entanglement in QM}
              
1. Being a considerable progress of the famous Aspect experiment [15], Weihs
experiment [5] for the first time fully enforced the condition of locality.
The spacelike separation of two ``observers" (Alice and Bob) is achieved by
sufficient physical distance (400m) between them and by ultrafast (the
duration of an individual measurement $<100$ns which is far less
than $1.3\mu s=400m/c$) and random setting of the analyzers. So the
possibility of any signal connecting Alice and Bob with velocity less than
or equal to the speed of light was certainly excluded.

The generalized Bell's inequality reads
\begin{equation}
S(\alpha,\alpha',\beta,\beta')=|E(\alpha,\beta)-E(\alpha',\beta)|
+|E(\alpha,\beta')+E(\alpha',\beta')|\leq 2
\end{equation}
where $E(\alpha,\beta)$ is the expectation value of two-photon correlation
with $\alpha$ and $\beta$ being the
directions of polarization analyzers of Alice and Bob. On the other hand,
the prediction of QM is 
\begin{equation}
S^{QM}_{max}=S^{QM}(0^\circ,45^\circ,22.5^\circ,67.5^\circ)=2\sqrt{2}=2.82>2.
\end{equation}
The 14700 coincidence events collected in 10s yield 
\begin{equation}
S^{exp}=2.73\pm 0.02
\end{equation}
which corresponds to a violation of inequality (7) of 30 standard deviation
and so strongly supports QM.

2. Being a remarkable realization of Franson's prominent proposal [16],
Tittel experiment [6] demonstrated the quantum correlation between two
(energy and time) entangled photons can be maintained over long distance
($>10$km). The coincidence counts between two interferometers were
fitted to the probability function as 
\begin{equation}
P=\frac{1}{4}\left( 1+Vexp\left\{ -\left[
\frac{\lambda(\delta_1-\delta_2)}{2\pi L_c}\right] ^2\right\}
\cos(\delta_1+\delta_2)\right)
\end{equation}
where $\delta_1$ or $\delta_2$ is the variable phase-difference in
either interferometer caused by the path length difference, $\lambda=1310$nm
is the wavelength of photon while $Lc$ is the single-photon coherence
length. The coefficient $V$ is called as the ``visibility",
$V\leq\frac{1}{\sqrt{2}}\cong 0.71$ inferred by the
Bell-inequality. But the experimental data showed $V^{exp}=81.6\pm 1.1\%>0.71$,
a violation of the
Bell-inequality by 10 standard deviation and a further strong support to the
QM.

3. A beautiful EPR experiment was performed by CPLEAR Collaboration at CERN
on $K^0\bar{K}^0$ system [7]. Alice and Bob were located at left and right
side with
distance $\sim 10$cm between. According to the prediction of QM, the
wave function of $K^0\bar{K}^0$ system is entangled as follows:
\begin{equation}
|\bar{\Psi}(t_a,t_b)\rangle=\frac{1}{\sqrt{2}}\left[
|K_S(t_a)\rangle_a|K_L(t_b)\rangle_b-|K_L(t_a)\rangle_a|K_S(t_b)\rangle_b
\right]
\end{equation}
where $t_a$ and $t_b$ are the proper time records at Alice and Bob sides while 
\begin{equation}
|K_S\rangle=\frac{1}{\sqrt{2}}\left[ |K^0\rangle+|\bar{K}^0\rangle\right] ,
\hskip 0.3in
|K_L\rangle=\frac{1}{\sqrt{2}}\left[ |K^0\rangle-|\bar{K}^0\rangle\right ].
\end{equation}
The ``asymmetry" is defined as
\begin{equation}
A(t_a,t_b)=\frac{I_{unlike}(t_a,t_b)-I_{like}(t_a,t_b)}
{I_{unlike}(t_a,t_b)+I_{like}(t_a,t_b)}
\end{equation}
where $I_{unlike}$ ($I_{like}$) is the intensity of event with $K^0\bar{K}^0)$
or $\bar{K}^0K^0$ ($K^0K^0$ or $\bar{K}^0\bar{K}^0$) detected. By
contrast, if the wave function is factorized or separable, i.e., only one
term is left in the expression (11), then the asymmetry would always be zero,
$A=0$. The experiment showed the value of $A(\Delta l)$ with $\Delta l$
(in cm) being the flight path difference:
\begin{equation}
A^{exp}(0)=0.81\pm 0.17, \hskip 0.3in A^{exp}(5)=0.48\pm 0.12
\end{equation}
in comparing with the prediction of QM:
\begin{equation}
A^{QM}(0)=0.93, \hskip 0.3in A^{QM}(5)=0.56.
\end{equation}
Thus the separability hypothesis is excluded with a confidence level
$CL>99.99\%$ and proves once again the validity of QM.

In all these EPR experiments mentioned above the entangled state i.e.,
two-particle state with quantum correlation over long distance, exhibits its
subtlety. For example, for the $K^0\bar{K}^0$ system described by Eq. (11),
only after Alice finds a $K^0$ (or $\bar{K}^0$) in the measurement at time
$t_a$, can Alice predict with $100\%$
certainty that Bob must finds a $\bar{K}^0$ (or $K^0$) at the same time
($t_a=t_b$). Since they
are separated over long distance, (10cm in CPLEAR experiment and even 
$>10$km in Tittel experiment), no information can be communicated
between them with a velocity equal to or less than the speed of light. In
other words, no local hidden variable (LHV) can exist as inferred by the
violation of Bell inequality. Therefore, the sudden nonlocal collapse of
wave function of entangled two-particle state (into a measured distinct
particle at Alice side with another one at Bob side) seems to be caused via
some ``spooky action at a distance" by Einstein, (see [17]).

\vskip 24pt
\noindent
{\bf III. Quantum state and wave function}

Let us try to understand the mystery posed by the experiments discussed
above, at least to some extent. Then it seems to us that the fundamental
interpretation of QM is involved.

In Dirac notation, a quantum state, e.g., a one-particle state in
one-dimensional space is denoted by an abstract state vector $|\Psi\rangle$
in Hilbert
space. In Hersenberg picture, there is no description either $x$ or $t$ in
$|\Psi\rangle$.
Only after some representation is chosen, can it get some description. For
instance, if we choose the eigenvector of the position $x$, $|x,t\rangle$,
as the base
vector and take the contraction (projection) of $|\Psi\rangle$ with
$|x,t\rangle$, we obtain the wave
function in configuration space:
\begin{equation}
\psi(x,t)=\langle x, t|\Psi\rangle.
\end{equation}
Alternatively, we can choose the eigenvector of momentum $p$, $|p,t\rangle$,
as the base
vector to get the wave function in momentum space ($p$ representation) as 
\begin{equation}
\varphi(p,t)=\langle p, t|\Psi\rangle.
\end{equation}
The two kinds of wave function, (16) and (17), are two different
descriptions for the same quantum state $|\Psi\rangle$. No one in the two is
more fundamental than the another one.

The wave functions in QM are not observable. But they are very useful in
linking the even more abstract state vector, say $|\Psi\rangle$, to the
potential possible
outcome in experiments if the latter are really performed on the state. For
example, we are going to measure the position $x$ of the particle, so we
choose $|x,t\rangle$ to characterize (represent) the ``apparatus" for $x$
measurement and
write down the wave function $\Psi(x,t)$. Note that, however [18],

(a) What contained in the $\Psi(x,t)$ is merely a ``fictitious measurement".
So being a
``probability amplitude", the wave function always contains the imaginary
number unit $i=\sqrt{-1}$ which is unobservable.

(b) According to the statistical interpretation by M. Born, $|\Psi(x,t)|^2$
is the
probability of finding the particle at position $x$ during the measurement
rather than that of the appearance of the particle before the measurement.

(c) Some times it was tacitly assumed that $x$ in the wave function is the
position coordinate of ``point particle". We don't think so. Instead, we
prefer to think that in the $1$S state of a Hydrogenlike atom the electron has
a spatial extension with radius $a/Z$ ($a=0.529\times 10^{-10}$m
being the Bohr radius and $Z$ being the charge number of nucleus). On the
other hand, when an electron is under high-energy collision, its spatial
extension may be compressed into a tiny one, say less than $10^{-18}$m [19].

\vskip 24pt
\noindent
{\bf IV. The relation between individual and its environment}

The existence state of any individual particle is depending on its
environment. This can be seen most clearly in the lifetime ($\tau$) of
an unstable particle. For instance, see the $\beta$-decay of nuclei. A
free neutron has $\tau=14.8$ minutes, while the $\tau$ of nuclide $^{11}_3Li_8$
is shortened to only 8.5$\mu s$. On the other hand, the $\tau$ of nuclide
$^{128}_{52}Te$ is extremely long: $\tau=2.1\times 10^{24}$ years. Many
nuclides, including the neutron in the neutron star (pulsar),
are stable against $\beta$-decay, i.e., they have $\tau=\infty$. This shows
that the lifetime of a neutron is strongly
influenced by its (nuclear) environment. Actually, it nearly has no
intrinsic stability. This can also be seen from the decay law:
\begin{equation}
N(t)=N_0e^{-t/\tau}, \hskip 0.3in
-\frac{dN/dt}{N(t)}=\frac{1}{\tau}=\lambda=const
\end{equation}
which means that a neutron at any time, as long as it has not decayed, has a
definite decay probability independent of its existing time already. Just
like the words said by the Chinese philosopher Zhuangzi (369BC-286BC):
``Just was born just died, just died just was born".

The mass of a particle is also depending on its environment. In the language
of quantum field theory, mass generation is only possible after the vacuum
undergoes a phase transition ([20], see also the explanation by Wilczek
[21], he speculated the discovery of Higgs particle in the years to come. We
had calculated the Higgs mass to be 138GeV [22]).

Now the nonlocal two-particle state in the EPR experiments is non-separable
before the measurement. As shown in Eqs. (11) and (12), the kaon at Alice (or
Bob) side is not either $K^0$ or $\bar{K}^0$, but neither $K^0$ nor $\bar{K}^0$.
The information that Alice
finds a $K^0$ (or $\bar{K}^0$) while Bob finds a $\bar{K}^0$ (or $K^0$) is
just created by them via
measurement at the same time.

The strong correlation among many particles is clearly shown in FQHE. The
ground state of $N$ electrons is described by the Laughlin wave function [23].
Every electron loses its independent feature. So the whole system exhibits
itself as an imcompressible fluid and the elementary excitation above the
ground state is a quasi-particle with fractional charge (say $e/3$) [2,3] and
carrying non-local information (such as an invisible string) [24]. Only
after we destroy the quantum coherence of FQHE, can an electron appear. 

Now we understand why the quark with fractional charge can not be deconfined
from a hadron (say a proton). This is because quark is not a particle in
the common sense, i.e., not a ``building block". The latter is only well
defined when it can be separated from its environment with the binding
energy $B$ much less than its rest energy $E_0$: $B/E_0\ll 1$. In fact, every
particle is changing during its separating process. When
we wish to pick a $u$ quark out of a proton, both this $u$ quark and other two
quarks ($u$ and $d$) are changed to such an extent that the whole proton is
destroyed and what we can see are other particles. 

The three valence quarks ($uud$) are suitable for describing the property of
a proton near its ground state. However, when the proton is under
high-energy collision, it would be better to use the parton model, i.e., to
resort to the picture of many sea quarks and gluons besides the valence
quarks. A proton is infinite in essence, it has various aspects in various
experiments. Not only dynamics, but also its ingredients are depending on
the character of experiment, i.e., on what we are looking for [18].

From individual particle to the whole environment, we see that they are all
infinite and mutually related. We tend to share the view point of Zurek
{\it et al} [25] that the environment plays a crucial role in destroying the
quantum
coherence and bringing the measuring process to an end. The apparatus, being
a part of environment, substitutes the original quantum correlation
(entanglement) in the measured object by the new entanglement between the
apparatus and the object, as shown in the microwave measurement of DNR
experiment [4]. The final stage of measurement is achieved at the screen or
detector (they are also part of the environment), where the wave packet of
particle is collapsed.

\vskip 24pt
\noindent
{\bf V. Is the moon there when nobody looks [17]?}

In the DNR experiment [4], the center-of-mass motion of atom is described by
the plane wave function, which served as a ``guiding field" for atom
motion. We can not think of a atom like a small ball with radius 0.1nm
passing through the double-slit with spacing $d=1.3\mu m$. Rather, we
should think of the atom like a wave packet with spatial extension exceeding
$d$. What we can do is discussing the wave function and its interference. The
particle feature of atom is displayed only at the final stage. Similarly,
the two-photon entangled state in Tittlel experiment [6] is correlated in
long distance ($>10$km). To talk about one photon being here or
there is meaningless since the single-photon coherence length $L_C$ is only
$10.2\mu m$. The CPLEAR experiment clearly shows that the entangled
state of $K^0\bar{K}^0$ system has a spatial extension $\sim 10$cm, far
exceeding the radius or Compton wave-length of a single kaon.

Hence, in our point of view, the so called ``wave-particle duality" means
the following. Before the quantum coherence of the motion of a particle is
destroyed by the measurement, we should handle it as ``wave" by Schrodinger
equation theoretically until it is detected and then shows its ``particle"
feature. This is a problem of different temperaments at two levels, not at
the same level.

We are now in a position to try to answer Einstein's question: ``What is the
physical reality?" It seems to us that a ``thing" should be defined at two
levels. An object when it is independent of the consciousness of mankind and
before the measurement is made, could be called as ``thing in itself". It is
something absolute in nature and containing no information. In QM, it is
denoted by a quantum state $|\Psi\rangle$ separated approximately from its
environment.
Then after some measurement is performed, it is turned into ``thing for us",
reflecting a series of experimental data. It is then something relative in
nature. Sometimes, we call it ``phenomenon". As J.A. Wheeler said: ``No
phenomenon is phenomenon until it is an observed phenomenon." The wave
function in QM is just playing the role for connecting the two levels of
matter via the fictitious measurement. In some sense, we may also claim
that: ``We can only see what we intend to see." Eventually, we will be
convinced by the Chinese saying: ``Oneness of heaven and man."

\vskip 24pt
\noindent
{\bf Acknowledgements}

The author wishes to thank P-z Bi, Y-z Chen, X-c Gao, Y-k
Huo, Z-d Liu, F-q Lu, Z-y Shen, J-y Tang, Y-s Wang, J-b Xu, S-q Ying, and
C-y Zhou for discussions. This work was supported in part by the NSF of
China .

\vskip 24pt
\noindent
{\bf References}

[1] D.C. Tsui, H.I. Stormer and A.C. Gossard, Phys. Rev. Lett. 48, 1559-1562
(1982).

[2] R. de-Picciotto, et al. Nature 389, 162-164 (1997).

[3] L. Saminadayar and D.C. Glattli, Phys. Rev. Lett. 79, 2526-2529 (1997).

[4] S. Durr, T. Nonn and G. Rempe, Nature 395, 33-37 (1998); Phys. Rev. Lett.
81, 5705-5709 (1998).

[5] G. Weihs, T. Jennewein, C. Simon, H. Weinfurter and A. Zeilinger, Phys.
Rev. Lett. 81, 5039-5043 (1998).

[6] W. Tittel, et al. Phys. Rev A. 57, 3229-3232 (1998); W. Tittel, J.
Brendel, H. Zbinden and N. Gisin, Phys. Rev. Lett. 81, 3563-3566 (1998).

[7] CPLEAR collaboration, A. Apostolakis, et al. Phys. Lett. B 422, 339-348
(1998).

[8] CPLEAR collaboration, A. Angelopoulos, et al. CERN-EP/98 -153. Oct. 7,
1998, Submitted to Phys. Lett B.

[9] M.O. Scully, B-G. Englert and H. Walther, Nature 351, 111-116 (1991).

[10] R. Feynman, R. Leighton and M. Sands, The Feynman Lectures on physics
Vol. III (Addison Wesley, Reading, 1965).

[11] P. Storey, S. Tan, M. Collett and D. Walls, Nature 367, 626-628 (1994);
Nature 375, 368 (1995).

[12] B-G. Englert, M.O. Scully and H. Walther, Nature 375, 367-368 (1995).

[13] H. Wiseman and F. Harrsison, Natrue 377, 584 (1995).

[14] G-j. Ni, Y-s.Wang, J-h. Qian and X-m. Fang, Physics changing the world,
(Press of Fudan University, 1998), Chapt 13.

[15] A. Aspect, J. Dalibard and G. Roger, Phys. Rev. Lett. 49, 1804-1807
(1982).

[16] J. D. Franson, Phys. Rev. Lett, 62, 2205-2208 (1989).

[17] N. D. Mermin, Phys. Today, April, 38-47 (1985).

[18] G-J. Ni, Acta Photonica Sinica 28, 112-117 (1999). Internet,
quantum-ph/9803001.

[19] G-j. Ni, Kexue (Science) 50 (2), 38-42 (1998).Internet,
quant-ph/9804013; to be published in a book ''Photon: old problem in light
of new ideas'', edited by V. Dvoeglazov, (Nova Science Publisher, 1999).

[20] G-j. Ni and S-q. Chen, Acta Physica Sinica (Overseas Edition), 7,
401-413 (1998).

[21] F. Wilczek, New Scientist, 10 April 1999, 22-37.

[22] G-j. Ni, S-y. Lou, W-f. Lu and J-f. Yang, Science in China (series A).
41, 1206-1215 (1998).

[23] R.B. Laughlin, Phys. Rev. Lett. 50, 1395-1398 (1982).

[24] S.Kivelson and M. Rocek, Phys. Lett. 156 B. 85-88 (1985).

[25] W. H. Zurek, Phys. Today, Oct. 1991, 36-44. Comments and Reply, ibid,
April 1993, 13-15, 81-90.

\vskip 24pt
\noindent
{\bf Figure Caption}

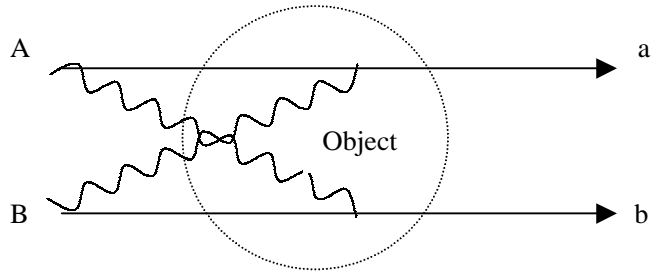
Figure 1: The measurement $A$ ($B$), being an operation (denoted by the arrow)
imposing on the object (denoted by dashed-line circle), creates the data
$a$ ($b$). If $A$ is not in conformity with $B$, then $A$ ($B$) becomes the
disturbance to $b$ ($a$) which is denoted by the wavy line.

\end{document}